# The AI-Native Software Development Lifecycle: A Theoretical and Practical New Methodology

Cory Hymel

Crowdbotics, cory.hymel@crowdbotics.com

*Abstract* - As AI continues to advance and impact every phase of the software development lifecycle (SDLC), a need for a new way of building software will emerge. By analyzing the factors that influence the current state of the SDLC and how those will change with AI we propose a new model of development. This white paper proposes the emergence of a fully AI-native SDLC, where AI is integrated seamlessly into every phase of development, from planning to deployment. We introduce the V-Bounce model, an adaptation of the traditional V-model that incorporates AI from end to end. The V-Bounce model leverages AI to dramatically reduce time spent in implementation phases, shifting emphasis towards requirements gathering, architecture design, and continuous validation. This model redefines the role of humans from primary implementers to primarily validators and verifiers — with AI acting as an implementation engine.

*Index Terms* - AI, SDLC, Software Development

## INTRODUCTION

The growing capabilities of artificial intelligence (AI) models, particularly large language models, have the potential to change the current software development lifecycle (SDLC) status quo and pave the way for a fully AI-native SDLC — an SDLC with AI-integrated end-to-end redefining roles and responsibilities. Recent advancements in AI model performance, driven by scaling up model size, dataset size, and compute power [1] have demonstrated remarkable progress across a wide range of cognitive tasks. Combined with the increasing multimodal capabilities, AI is rapidly approaching human-level performance in various domains specific to the SDLC, such as requirements engineering, resource allocation, budget estimations, code generation, code suggestion, automated testing, and others.

## II. What It Means to Be AI "Native"

The term "native" is a bit ambiguous. We've heard "mobile native," "web-native," and now "AI-native." The term "native" is often used as a prefix to a company, system, application, or methodology. Regardless of which entity it modifies, the high-level concept of "native" remains consistent: it involves the pervasive use of a tool, technology, or methodology and the necessary infrastructure in all sub-components of an entity, rather than merely adding it to an existing non-native-based entity [2].

When we use the term AI-Native SDLC, we mean a software development lifecycle where AI is seamlessly integrated into every phase — from planning to maintenance — enhancing efficiency, accuracy, and innovation across the entire process.

## III. The Potential for AI to Transform the Software Development Lifecycle (SDLC)

While AI has the ability to disrupt, displace, or replace discrete functions of the SDLC, we want to examine its impact on the SDLC as a whole. The current SDLC is built around roles and responsibilities that are not able to cross into other domains due to human limitations and lack of multidisciplinary expertise. AI introduces the ability to extend an individual's skills and knowledge beyond traditional means— creating a window of opportunity to disrupt the SDLC.

**Planning**: AI-powered project management tools can help to predict potential bottlenecks and resource constraints early, which helps with more proactive decision-making.

**Design:** By analyzing user interaction data, AI can suggest design improvements that enhance user experience and bridge the gap between user research and design implementation.

**Development**: AI generates code snippets but it can also review and optimize code for performance and security in real-time. AI-driven automation creates comprehensive test cases and simulates scenarios reducing the need for separate quality assurance stage, ensuring a more robust validation of the software.

**Maintenance**: AI can predict and prevent potential failures by analyzing usage patterns and system performance metrics — ensuring higher reliability and uptime.

Integrating these capabilities and context across all phases, AI has the potential to transform the SDLC from a series of isolated tasks into an intelligently integrated, continuous, and highly efficient process. This holistic transformation can lead to shorter development cycles, reduced costs, and higher-quality software. As recently summarized by Smit et al. 2024 [3]: Over the past years, different possibilities of application have evolved. One example is the concept of Software 2.0 [4]. Contrary to Software 1.0, where developers explicitly wrote code to define behavior, Software 2.0 uses neural networks to model the behavior required [5].

## The Current State of AI in the SDLC

The growing capabilities of artificial intelligence (AI) models, particularly large language models, have the potential to fundamentally disrupt the software development lifecycle (SDLC) and pave the way for a fully AI-native SDLC – an SDLC with AI integrated end-to-end redefining roles and responsibilities. Recent advancements in AI model performance, driven by scaling up model size, dataset size, and compute power [1] have demonstrated remarkable progress across a wide range of cognitive tasks. Combined with the increasing multimodal capabilities, AI is rapidly approaching human-level performance in various domains specific to the SDLC, such as requirements engineering, resource allocation, budget estimations, code generation, code suggestion, automated testing, and others.

*II. Planning and Requirements Gathering*

The planning and requirements phase of any software project, regardless of the model being followed, is incredibly important. There are varying reports of project failure due directly to bad requirements: Hall et al. [6] report that a large proportion (48%) of development problems stem from problems with the requirements, The Standish Group 44% of the reasons for failed projects have their origin in bad requirements gathering [7], CIO Analyst report 71% [8].
This phase largely consists of taking natural language requirements and translating them into discrete tasks. A recent study found that 45% of companies surveyed were already using general LLM tools such as ChatGPT to help streamline the planning and requirements phase [9].

Sultanov et al. [10] show that using reinforcement learning models are able to understand different levels of requirements abstraction. This approach is an example of how models can automatically generate traceability between high-level and low level requirements.

*III. Design and Architecture*

Design and architecture, in most cases, are separate efforts that require separate skill sets to complete. Design teams take the requirements and translate them into UI/UX assets and flows that are validated by customers. Architecture teams create the software architecture needed to support both functional and non-functional requirements. What's interesting is that when designing to support both of these, decisions made by the design teams can have a significant impact on the underlying technology chosen to support them.

*IV. Implementation and Coding*

The integration of AI into the coding process has already created a shift in software development practices, largely in part to the efficiency gains it gives developers. A large-scale study of GitHub Copilot users found that developers accepting AI suggestions completed their tasks 55.8% faster on average compared to those not using AI assistance [11]. In an internal study from Accenture involving around 450 developers, the use of Copilot resulted in more throughput and higher quality from their developer teams [12]. It's also been found that for 80% of developers who use AI code assistants, their overall experience as a software developer improved [11].

*V. Testing and Quality Assurance*

A critical stage of the software development life cycle (SDLC) is software testing, which looks for software systems' flaws, mistakes, or vulnerabilities. Conventional testing approaches rely significantly on physical labor, which can be expensive, time-consuming, and prone to errors [13]. Software testing is a field experiencing rapid innovation thanks to AI techniques like automation and machine learning, transforming quality assurance procedures [14]. As AI code generation becomes more prevalent the need for increased rigor of testing and quality assurance testing is critical with roughly 50% of organizations reporting at least one security incident over a period of 12 months [15]. Today, there are already a number of tools being leveraged to help this phase of the SDLC. Recent studies indicate that the efficiency in generation of test suites through ChatGPT-based system exceeds 70% [16]. This enhanced coverage translates to earlier detection of potential bugs and improved overall software quality.

*VI. Deployment and Maintenance*

Deployment is the last three feet of any software project but can also at times be the most cumbersome. AI-driven continuous integration and continuous deployment (CI/CD) is a software engineering methodology that can help software developers create, test, and deploy applications faster and with greater efficiency [17]. It involves automating the process of integrating code changes, testing and releasing software updates, and automating deployments to ensure that the latest changes are live in production as soon as possible [17]. In the maintenance phase, AI is proving invaluable for predictive maintenance and performance optimization. AI systems' ability to disparate data sets, such as log data, performance metrics, and user behavior patterns, allow them to identify potential problems and suggest optimizations. By being able to synthesize this data quickly, studies have found models' ability to identify and classify soft bugs has improved — with up to 86% precision [18].

*VI . Why We Need to Think Bigger*

Shafiq et al. summarize it well, "The software engineering (SE) community is continuously looking for better and more efficient ways of building high-quality software systems. However, in practice, the strong emphasis on time to market tends to ignore many well-known SE recommendations. That is, practitioners focus more on programming, as compared to requirements gathering, planning, specification, architecture, design, and documentation – all of which are ultimately known to greatly benefit the cost-effectiveness and quality of software systems." [19].

Many of the tools and techniques listed above show additive integration of AI into the SDLC but fail to fully rethink how a true, AI-native SDLC would function. In the next section we look to analyze key aspects of the SDLC and how they will be drastically changed if made AI-native.

**AI INFLUENCE ON THE SDLC**

It's hard to truly know what a full AI-native SDLC will be until it's here. When the mobile phone came out, no one would have imagined that you'd be using it to order on-demand rides through a dynamic network of on-demand drivers. It helps to analyze foundational ideas that could be impacted when AI becomes truly infused into every piece of the SDLC. Rather than looking at each direct phase of the existing SDLC, we analyze aspects that influence the current SDLC: Speed, Teams, Intelligence, Resources, and Demand.

*I . (Speed) End of 2-Week Sprints*
The current 2-week sprint allows for the minimum amount of time between management check-ins that gives engineers time to build enough to be worth checking in on. The Scrum framework, formalized by Ken Schwaber and Jeff Sutherland in the 1990s, introduced the concept of "sprints" - time-boxed iterations of typically 1 month or less. The original Scrum guidelines suggested a 30-day sprint as a productive balance between too much overhead (for very short sprints) and too much risk (for very long sprints).

As Scrum started gaining popularity in the early 2000s, many teams found 30-day sprints to be too long — making it difficult to adapt to changing requirements and get frequent feedback. Shorter sprint lengths of 2-4 weeks became more common. [20] [21]

The 2-week sprint cycle emerged as a popular "sweet spot" that provided several benefits:
- Long enough for teams to make meaningful progress.
- Frequent enough to make course corrections if needed.
- Short enough to maintain focus and avoid significant deviations from the sprint goal.

2-week sprints essentially consist of two types of work: engineering and reporting. Engineering work being the coding itself and reporting being all non-coding tasks such as milestone updates, storypoint updates, housekeeping items, etc.

With the efficiency gains of AI we've seen both of these areas have significant improvement gains:

- **Engineering**: AI coding assistants have been shown to increase developer speed by 50% or more. [22]

- **Reporting**: There have been many studies showing that ChatGPT substantially raises average productivity by both decreasing the time taken to complete reporting-related tasks and raising output quality [23] [24]. While there is not a consolidated statistic on "non-engineering tasks," 91% of the field expect AI to have at least a moderate impact on them, with 21% already using it today [25] with aggressive estimates that AI will replace 80% of today's project management activities by 2030 [26].

Taking this into account, development will be reduced by 50% and reporting tasks will be reduced by some significant amount, we may estimate up to 30%, then it's reasonable to expect that 2-week sprints will quickly become too long.

*II . (Teams) Humans as verifiers, not creators.*

There are a number of reasons why this shift will happen. One that's most concrete and historically significant is money. If we look today to build a business case on migration or assimilation of AI code generation we can calculate the output cost per day of a human software engineer (SWE) compared to an AI model.

TABLE I
DAILY COST OF SOFTWARE ENGINEER

| | |
|---|---|
| Software Engineer Salary: | $220,000 |
| Benefits, taxes, etc.: | $92,000 |
| Total:(time of writing) | $312,000 |
| Working days per year: | 260 |
| Cost per day for SWE: | $1,200 |
| Avg. # lines of code committed per day | 100 |
| **Cost per line of code** | **$12** |

On average, SWEs will commit roughly 100 lines of code per day.

TABLE 2
DAILY COST OF EQUIVALENT AI MODEL

| | |
|---|---|
| Avg. number of GPT-3 tokens per line | 10 |
| Avg. # lines of code committed per day | 100 |
| Price for GPT-3 | $0.02 / 1K TOKENS |
| Cost per day: (time of writing) | **$0.12** |
| Cost per line of code | **$0.002** |

As you can see, there's a significant difference in spend between the two models. There are a number of other benefits to consider with using a model to generate code as compared to a human: models don't take breaks, models don't quit, models retain context, models take the same amount of time to generate code whether it's a prototype or production-ready code, and models make mistakes — though they make them quickly, with the cost to change or rewrite being insignificant.

While we don't see models replacing software engineers today, the SDLC will change as the models continue to evolve. Sam Altman recently put it well, "you can assume that models are as good as they are going to get and design around that or you can assume they are going to get vastly better and design for that." Based on growth trends we can expect models to continue to improve to a point that replacing human engineers with models is not only viable from an output perspective but financially prudent.

*III. Teams of Tomorrow*

Software teams tomorrow will be "integrations of multiple GPT agents into a unified framework that enhances AI's problem-solving prowess. Our goal is a multi-GPT agent SE framework that streamlines software development, maintenance, bug detection, and documentation. The framework autonomously handles these tasks, improving efficiency and reducing development time" [27]. Humans will act as input vectors and validators of the outputs from the models and agents. As outlined in Figure 1, each agent also has an accompanying quality agent that acts as a checksum to the originator agent's output. This new composition puts heavy emphasis on validation and verification of the outputs.

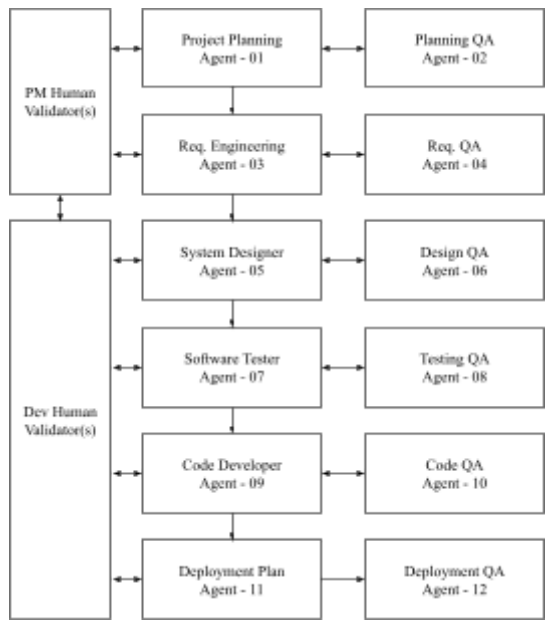

FIGURE I

Human-in-loop as validators for development teams of the future. Adaptation from [27].

*Example*: "The primary role of Agent-01 is to define and lay out the entire project's scope, objectives, and the steps required to achieve those objectives. Agent-02, responsible for project planning quality analysis, focuses on ensuring the quality and effectiveness of the project plan created by Agent-01. Agent-02 assesses whether the project plan is comprehensive, feasible, and aligns with the organization's goals" [27]. We've seen this new team structure already become a reality with tools like MetaGPT that organize a multi-agent system within the context of SOPs [28].

*IV. (Intelligence) AI as a knowledge management*

Software engineering (and the SDLC) involves a multitude of knowledge-intensive tasks: analyzing user requirements for new software systems; identifying and applying best software development practices; collecting experience about project planning and risk management; and many others (Birk, et al., [29]).

Knowledge management (KM) is seen as a strategy that creates, acquires, transfers, brings to the surface, consolidates, distills, promotes creation, sharing, and enhances the use of knowledge in order to: improve organizational performance; support organizational adaptation, survival, and competence; gain competitive advantage and customer commitment; improve employees' comprehension; protect intellectual assets; enhance decisions, services, and products; and reflect new knowledge and insights [30].

AI is uniquely positioned to be a transformative technology by centralizing KM and making it not only more accessible but also extensible as it:
- Automatically captures knowledge from various sources (code repositories, documentation, communication channels) without manual intervention.
- Organizes and connects disparate pieces of information to create a comprehensive knowledge graph of the entire SDLC process.
- Provides contextually relevant information to team members based on their current task, role, and project phase.
- Identifies knowledge gaps and suggest areas where more documentation or clarification is needed.
- Learns from past projects to provide insights and recommendations for current and future projects.
- Facilitates natural language queries, allowing team members to ask complex questions and receive accurate, context-aware responses.

*V. (Resources) Global talent pool, Continuous delivery.*

The concept of global software development has long been a tantalizing prospect for the industry, promising around-the-clock productivity and access to a diverse, worldwide talent pool. One of the most prominent attempts to realize this vision was the Follow-The-Sun (FTS) delivery model. FTS can be explained as "Hand-off work every day from one site to the next as the world turns (USA to India, for example). Thus, reduce the development duration by 50% if there are two sites and by 67% if there are three sites." [31].

This approach aimed to optimize productivity by distributing tasks across three global teams, enabling continuous development as geographical teams handed off work in alignment with time zone changes. The FTS model, in theory, offered several advantages:

- Continuous 24-hour development cycles
- Faster time-to-market for products
- Access to a global talent pool
- Potential cost savings from labor arbitrage

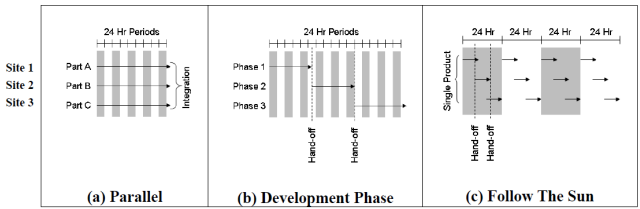

FIGURE II

FTS COMPARED TO OTHER GLOBALLY DISTRIBUTED CONFIGURATIONS. WE REFER TO PARALLEL AND PHASE-BASED AS "CONVENTIONAL GLOBAL CONFIGURATIONS." [32]

Despite its promising concept, the FTS model faced significant challenges that led to its limited adoption and success:

**Context Transfer**: The "heavy tax of context transfer" was a critical bottleneck. Teams struggled to effectively communicate project status, ongoing issues, and critical decisions across time zones and cultures.

**Coordination Costs**: The overhead of managing distributed teams often outweighed the benefits of continuous development.

**Cultural and Language Barriers**: Miscommunications and misunderstandings were common, leading to errors and delays.

**Inconsistent Development Practices**: Maintaining consistent coding standards and practices across geographically dispersed teams proved challenging.

All of the challenges, however, are uniquely solvable with the use of AI.

**Context Transfer**: AI can serve as a central, always-available knowledge repository. It can capture, organize, and disseminate project context, decisions, and status updates in real time, ensuring all team members have access to the latest information regardless of their time zone (as expanded on in the previous section).

**Coordination**: AI-powered project management tools can automate task allocation, progress tracking, and resource management across distributed teams, significantly reducing coordination overhead.

**Cultural and Language Barriers**: Advanced Natural Language Processing (NLP) models can provide real-time translation and cultural context, facilitating smoother communication between diverse team members.

**Consistent Practices:** AI can enforce coding standards by automatically reviewing code for consistency and even generating code that adheres to established team practices (such as OpenAI Codex, Amazon CodeWhisperer, and others).

**Asynchronous Collaboration**: LLMs acting as knowledge management agents can facilitate more effective asynchronous communication, reducing the need for real-time meetings and mitigating time zone fatigue as well as reducing overall handoff times between FTS teams.

As we've outlined, AI's ability to act as a comprehensive knowledge management tool is the key to enabling true global delivery. By capturing, organizing, and distributing knowledge seamlessly across time zones and cultures, AI can create a shared, always-up-to-date context for all team members. This unlock will allow for true continuous, global development, accelerating the rate at which software is created.

*VI. (Demand) Jevons Paradox*

The Jevons Paradox in energy economics, posits that as artificial intelligence (AI) significantly reduces the cost and effort of software development, it will paradoxically lead to an increased demand for software, rather than a reduction in overall development activity. This behavior is rooted in the economic principle that as a resource becomes more efficient to use, its consumption often increases [40].

In the context of software development, AI will disrupt the current status quo, which will lead to a surge in demand for new software applications. This will (*Speed*) increase the speed at which new software can be created, (*Teams*) reduce the cost of software development, (*Intelligence*) expand software into new domains, and (*Resources*) expand the eligible talent pool. This disruption to the current status quo will lead to a heightened demand for new software applications. To meet this exponential growth in demand, the software industry will need to not only adopt AI tooling but also re-think how the software development life cycle is structured.

### New Approach: V-Bounce

Traditional agile methodologies, while effective in their time, may no longer be sufficient to fully leverage the potential of AI in software development. As Ziegler et al. note, the integration of AI technologies into software engineering practices is leading to a paradigm shift in how software is conceived, designed, and implemented [33]. This shift calls for a new approach that can accommodate the speed, efficiency, and capabilities that AI brings to the development process.

We propose a new approach to the SDLC, one that is built on three key assumptions, derived from current trends and research in AI-assisted software development:

1. Code generation becomes near-instantaneous and cost-effective: Recent advancements in large language models have demonstrated the ability to generate high-quality code rapidly and accurately [34].
2. Natural language becomes the primary programming interface: Studies have shown increasing success in translating natural language descriptions into executable code, suggesting a future where programming could be primarily driven by natural language inputs [35].
3. Human roles shift from creators to verifiers: As AI takes on more of the code generation tasks, human developers' roles are likely to evolve towards verification, high-level design, and strategic decision-making [36].

Based on these assumptions, we propose an adaptation of the V-model, commonly used in systems engineering and mechatronic product development, to create a "V-Bounce" model for AI-native software development. This model aims to combine the rigorous validation and verification processes of the V-model with the flexibility and iterative nature of agile methodologies, all enhanced by AI capabilities. "Bounce" comes from the limited amount of time meant to be spent at the bottom of the "V."

*I. V-Model Overview*
While there is competing existing literature on who came up with the V-model first, the general consensus is that it was developed simultaneously and independently in Germany and the United States in the late 1980s. The model emphasizes the importance of validation and verification throughout the development process, representing these activities in a V-shaped diagram. The V-model was, and continues to be, used heavily in mechatronic product development, where it has been particularly valuable due to its ability to handle the complexities of integrating mechanical, electrical, and software components.

The IAPM summarizes the V-model as follows: "The V-model can basically be divided into three phases. In the verification phase, the requirements are recorded and converted into a system analysis. These requirements become increasingly specific according to the top-down principle and represent the left-hand side of the V-model. In the second phase, the coding phase, the product is developed. This is followed by the validation phase, in which the bottom-up principle is applied and tested. This is the right-hand side of the V. You start with the tests at the unit level and work your way up to the tests at the system level. This phase and the entire project end with product acceptance." [37].

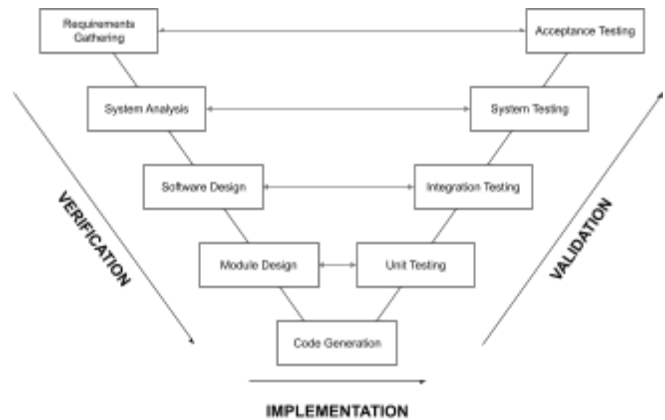

Figure III
Overview of the V-model

*II. Principles of the V-Model*

The V-Model provides a great foundation for our three assumptions: code will be instantaneous and nearly free; natural language is the primary means of programming; and humans are verifiers rather than core creators in the SDLC. We also find it applicable because of the heavy testing of machine-generated code needed, which naturally shifts humans into the verifier role. It's important to note that the core principles of the existing V-Model have been in use for over two decades. These are:

**Continuous Testing Integration**: Incorporation of testing activities throughout the entire development process, from initial requirements gathering to final deployment. This ensures ongoing, iterative quality assurance rather than relegating testing to a final phase.

**Parallel Planning of Development and Testing**: Each development stage in the V-Model is associated with a corresponding testing phase. This parallel structure enforces concurrent planning of both development and testing activities and forces validation forethought throughout the process.

**Precise Requirements Definition:** A fundamental tenet of the V-Model is the establishment of clear, concise, and unambiguous requirements. Precision is critical for developing effective test cases and ensuring the final product aligns with stakeholder expectations.

**Integration of Development and Testing Processes**: Rather than treating development and testing as separate entities the V-Model advocates for close integration. This principle promotes collaboration between developers and testers.

## V-MODEL INTO V-BOUNCE

The V-Bounce model, while based in the traditional V-Model framework, introduces significant adaptations to optimize for AI integrations throughout every phase of development. The largest change is time allocated across each phase and, most notably, the time spent in implementation (coding). With code generation being near instantaneous and free, teams spend minimal time at the bottom of the V — hence the term "bounce". However, there are opportunities for AI to be integrated throughout every phase of the process, significantly reducing overhead and speeding up sprint cycle time.

### I. AI Integration

The integration of AI throughout every phase of the existing V-Model is core to enabling V-Bounce. While the V-Model relies primarily on human expertise across all stages, the V-Bounce model incorporates AI from requirements analysis through to code generation, testing, and maintenance. Recent research has explored the integration of AI into the V-model for mechatronic product development. Nüßgen et al. conducted a comprehensive literature review on this topic, identifying potential AI integration points throughout the V-model [38] — see Appendix A for more examples.

### II. Time Allocation

A key distinction between the V-Model and V-Bounce is the time allocation across phases. The traditional V-Model typically dedicates a substantial portion of the development cycle to the implementation phase at the bottom of the V; the V-Bounce model drastically reduces the time spent in this phase. This compression is made possible by AI-driven code generation.

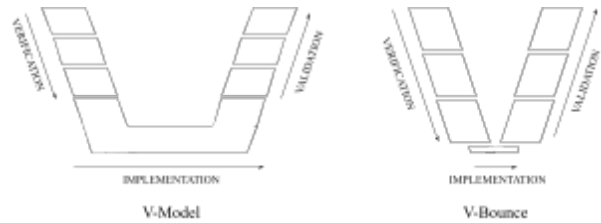

FIGURE IV
COMPARISON OF RELATIVE TIME SPENT PER PHASE V-MODEL VS. V-BOUNCE

Code generation, however, requires high levels of requirements decomposition in order to effectively generate accurate code. The V-Bounce model shifts emphasis to requirements, architecture, and design to ensure high quality input is given to code generation models.

### III. Test Creation

One of the hallmarks of the V-Model is the relationship between each development phase (**Verification**) and its corresponding testing phase (**Validation**). This has led critics to observe less flexibility in requirements changes, which can be time consuming and expensive, and is not ideal for complex projects.

The key differentiation in V-Bounce is leveraging AI to generate tests in real-time as requirements are being created. This simultaneous process turns weak areas of the previous V-Model into strengths.

**Continuous Test Creation**: As requirements are input (often in natural language), the AI system not only processes and formalizes these requirements but also immediately begins generating corresponding test cases. This ensures that each requirement is testable from the moment it's conceived. This continuous validation process significantly reduces the time gap between requirement definition and test creation.

**Early Detection of Ambiguities**: By attempting to create tests in real time, the AI can quickly identify ambiguities or inconsistencies in the requirements. This prompts immediate clarification, reducing the risk of misunderstandings later in the development process. For instance, if a requirement states: "The system should process orders quickly," the AI might flag this as ambiguous and prompt for clarification:
What is the definition of "quickly" in this context? Are there different processing time expectations for different types of orders? How should the system prioritize orders during high-volume periods?

**Traceability by Design**: Simultaneous generation of requirements and tests creates an inherent traceability matrix. Each test is directly linked to its corresponding requirement, facilitating easier tracking and impact analysis throughout the project lifecycle. If the requirement changes, the AI can immediately identify which tests are affected and need updating. This level of traceability helps manage complexity, making the V-Bounce model well-suited for projects that the traditional V-Model might struggle with.

**Adaptive Test Suite**: As requirements evolve or change, AI can dynamically update the test suite, ensuring that the testing strategy always aligns with the current state of requirements. If a requirement changes from "Users can upload images up to 5MB" to "Users can upload images up to 10MB", the AI can automatically: Update relevant test cases with the new file size limit, generate new edge cases (e.g., testing with 9.9MB files), flag any related requirements or tests that might be impacted by this change.

*VI. V-Bounce in Agile*

The V-Bounce model, as with the V-Model, can be easily adapted into agile development. In agile, V works as a V + V-centric approach — assuming that the V-model is not used like one large waterfall approach. Instead, the Agile V-model must ensure that the sprints themselves will become Vs according to the V-model.

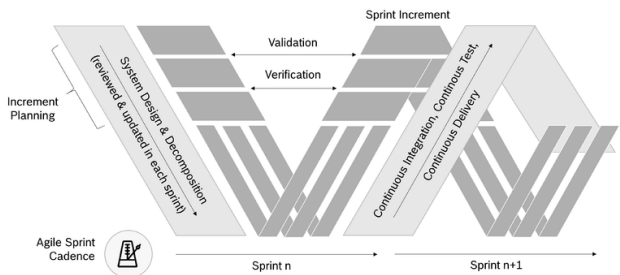

FIGURE V
AGILE V+V MODEL [39]

*VI. Anatomy of V-Bounce Phase*

Each phase in the V-Bounce model is an iterative process characterized with tight collaboration between humans and AI. Unlike traditional development models where humans perform most tasks, the V-Bounce phase leverages AI to handle the bulk of the complex, time-consuming work, while humans focus on verification of outputs. There are six main activities that occur within a V-Bounce phase as shown in Figure V.

**Input**: The phase begins with input that can come from either humans or artifacts from the previous phase. As an example, for the *Requirements* phase, input would be given in the form of natural language. In the *Implementation* phase, input would be the output artifacts of the previous phases.

**AI Generation**: The AI model processes the Input to generate outputs. This could be in the form of Product Requirement Docs in the *Requirements* phase, design mockups in the *Design* phase, or Code in the *Implementation* phase.

**Human Review**: This step is critical in that humans act as verifiers reviewing the AI-generated output, providing feedback, requesting modifications, or approving the results.

**Refinement**: Based on human feedback, the AI rapidly iterates and refines its output.

**Approval**: Once satisfied with the AI's output, humans perform a final verification, ensuring that the phase's objectives have been met and align with the overall project goals.

**Knowledge Capture**: Throughout the process, the AI system captures new knowledge and insights, which are stored and can be applied in future phases or projects.

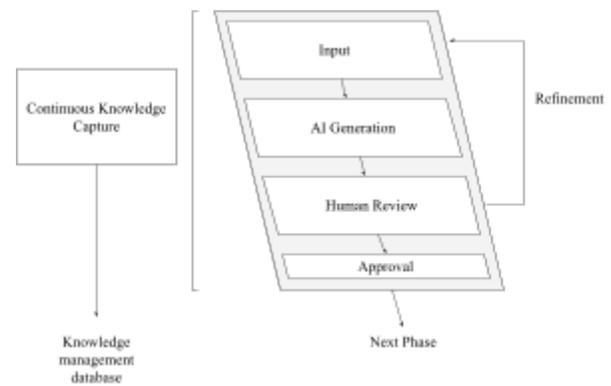

FIGURE VI
THE ANATOMY OF A V-BOUNCE PHASE

## CONCLUSION AND FUTURE RESEARCH OPPORTUNITIES

Implementation V-Bounce throughout the entire software development lifecycle as proposed in the paper, has yet to be done. However, integration of artificial intelligence into the software development life cycle has already begun. AI has already started to show signs that our assumptions that in the future —that code generation will become near-instantaneous and cost-effective, natural language will become the primary programming interface, and human roles will shift from creators to verifiers – are not far-fetched.

This whitepaper was meant to take a step back to look at AI's impact on the overall SDLC and propose that a new method is needed to facilitate the future of development with AI playing a central role throughout. While the V-Bounce model presents a framework for AI-native software development, its introduction also opens up several avenues for future research:

**Empirical Validation**: Large-scale studies are needed to quantify the benefits and challenges of implementing the V-Bounce model in various organizational contexts and project types.

**AI Reliability and Bias**: Further research is required to ensure the reliability of AI-generated code and tests — and to develop methods for detecting and mitigating potential biases in AI outputs.

**Ethical and Legal Frameworks**: As AI takes on a more prominent role in software development, research into ethical guidelines and legal frameworks for AI use in this context becomes crucial.

**AI-Driven Requirements Engineering**: Further investigation into how AI can improve the elicitation, analysis, and management of software requirements could significantly enhance the early stages of development.

**Adaptive AI Systems**: Research into AI systems that can learn and adapt from project to project, continuously improving their performance and accuracy in software development tasks.

**Security Implications**: Studies on the security implications of AI-generated code and how to ensure robust security measures in AI-native development processes are essential.

## Acknowledgements

This paper is funded and sponsored by Crowdbotics. Special thanks to Anand Kulkarni for feedback and for building a vision that led to the idea of V-Bounce. Special thanks to Darcy Jacobsen from The Wednesday Group for continuous proofing and editing before publication.

## Author Information

**Cory Hymel**, VP Research & Innovation, Crowdbotics.

## References

[1] J. Kaplan et al., Scaling Laws for Neural Language Models. 2020. [Online]. Available: https://arxiv.org/abs/2001.08361
[2] Ericsson, "Defining AI Native: A Key Enabler for Advanced Intelligent Telecom Networks," ericsson.com. https://www.ericsson.com/49341/assets/local/reports-papers/white-papers/ai-native.pdf (accessed June 4, 2024).
[3] Danie S. et al., "The impact of GitHub Copilot on developer productivity from a software engineering body of knowledge perspective" (2024). AMCIS 2024 Proceedings. 10. https://aisel.aisnet.org/amcis2024/ai_aa/ai_aa/10
[4] A. Karpathy, "Software 2.0," Available: https://karpathy.medium.com/software-2-0-a64152b37c35 (accessed June 11, 2024).
[5] Sheng, Y., Vo, N., Wendt, J. B., Tata, S., and Najork, M. (2020). "Migrating a Privacy-Safe InformationExtraction System to a Software 2.0 Design." in CIDR.
[6] Hall T, Beecham S, Rainer A, 2002, "Requirements problems in twelve software companies: an empirical analysis", Software IEEE Proceeding 149(5):153–160
[7] Project failure: Standish Group – Chaos report 1995, http://pmbullets.blogspot.in/2010/04/project-failurestandish-group-chaos.html (accessed July 7, 2014)
[8] Christopher Lindquist, "Fixing the Software Requirements Mess." cio.com. https://www.cio.com/article/255253/developer-fixing-the-software-requirements-mess.html (accessed July 15, 2024).
[9] Y. Zelianko, "The Transformative Impact of AI in Software Development: A Survey-Based Analysis." techreviewer.co. https://techreviewer.co/blog/the-transformative-impact-of-ai-in-software-development-a-survey-based-analysis (accessed July 10, 2024).
[10] H. Sultanov and J. H. Hayes, "Application of reinforcement learning to requirements engineering: requirements tracing," in 2013 21st IEEE International Requirements Engineering Conference (RE), 2013, pp. 52–61. doi: 10.1109/RE.2013.6636705.
[11] S. Peng, E. Kalliamvakou, P. Cihon, and M. Demirer, The Impact of AI on Developer Productivity: Evidence from GitHub Copilot. 2023. [Online]. Available: https://arxiv.org/abs/2302.06590
[12] Salva, R., Zhao, S., Kirschner, H., Androncik, A., and Schocke, D., *GitHub Copilot: the AI pair programmer for today and tomorrow*, (Nov 2023). Accessed June. 18, 2024. [Streaming Video]. Available: https://www.youtube.com/watch?v=AAT4zCfzsHI.
[13] K. Mullangi, S. S. Maddula, M. A. Shajahan, and A. K. Sandu, "Artificial Intelligence, Reciprocal Symmetry, and Customer Relationship Management: A Paradigm Shift in Business", Asian bus. rev., vol. 8, no. 3, pp. 183–190, Dec. 2018.
[14] A. Varghese and M. T. I. Bhuiyan, "Emerging Trends in Compressive Sensing for Efficient Signal Acquisition and Reconstruction", TMR, vol. 5, pp. 28–44, Feb. 2020, Accessed: June 24, 2024. [Online]. Available: https://upright.pub/index.php/tmr/article/view/119
[15] Code Intelligence, "The Future of Productive Development: Self-Learning AI for a Secure Tomorrow," Code Intelligence, Bonn, Germany. Accessed June 24, 2024. [Online]. Available: https://www.code-intelligence.com/ai.
[16] Y. Chen, Z. Hu, C. Zhi, J. Han, S. Deng, and J. Yin, ChatUniTest: A Framework for LLM-Based Test Generation. 2024. [Online]. Available: https://arxiv.org/abs/2305.04764
[17] A. S. Mohammed, V. Saddi, S. Gopal, D. Selvaraj, and M. Naruka, "AI-Driven Continuous Integration and Continuous Deployment in Software Engineering," Mar. 2024, pp. 531–536. doi: 10.1109/ICDT61202.2024.10489475.
[18] N. K. Nagwani and J. S. Suri, "An artificial intelligence framework on software bug triaging, technological evolution, and future challenges: A review," International Journal of Information Management Data Insights, vol. 3, no. 1, p. 100153, 2023, doi: https://doi.org/10.1016/j.jjimei.2022.100153.
[19] S. Shafiq, A. Mashkoor, C. Mayr-Dorn, and A. Egyed, "A Literature Review of Using Machine Learning in Software Development Life Cycle Stages," IEEE Access, vol. 9, pp. 140896–140920, 2021, doi: 10.1109/ACCESS.2021.3119746.
[20] "Why Are Sprints Two Weeks Long?" createfuture.com. https://createfuture.com/blog/why-are-sprints-two-weeks-long (accessed June 18, 2024).
[21] A. Jacob, "2 Week Sprints - the Power of Iterations." agiledad.com. www.agiledad.com/post/2018/12/15/2-week-sprints-the-power-of-iterations (accessed June 18, 2024).
[22] E. Kalliamvakou, "Research: quantifying GitHub Copilot's impact on developer productivity and happiness." github.blog. https://github.blog/news-insights/research/research-quantifying-github-copilots-impact-on-developer-productivity-and-happiness/ (accessed June 20, 2024).
[23] S. Noy and W. Zhang, 'Experimental evidence on the productivity effects of generative artificial intelligence,' Science, vol. 381, no. 6654, pp. 187–192, 2023.
[24] Md. M. I. Shamim, "Artificial Intelligence in Project Management: Enhancing Efficiency and Decision-Making," vol. 1, pp. 1–6, Apr. 2024, doi: 10.62304/ijmisds.v1i1.107.
[25] "Shaping the Future of Project Management With AI," pmi.org. https://www.pmi.org/learning/thought-leadership/ai-impact/shaping-the-future-of-project-management-with-ai (accessed June 18, 2024).
[26] "How AI Will Reinvent Program and Portfolio Management." gartner.com.

[26] https://www.gartner.com/en/newsroom/press-releases/2019-03-20-gartner-says-80-percent-of-today-s-project-management (accessed June 24, 2024).
[27] Z. Rasheed et al., Autonomous Agents in Software Development: A Vision Paper. 2023. [Online]. Available: https://arxiv.org/abs/2311.18440
[28] S. Hong et al., MetaGPT: Meta Programming for A Multi-Agent Collaborative Framework. 2023. [Online]. Available: https://arxiv.org/abs/2308.00352.
[29] Birk A., Surmann D., Althoff K. 1999, 'Applications of Knowledge Acqusition in Experimental Software Engineering', 11th European Workshop on Knowledge Acquisition, Modeling, and Management, pp.67-84.
[30] A. Alawneh, E. Hattab, and W. Al-Ahmad, "An Extended Knowledge Management Framework During the Software Development Life Cycle.," The International Technology Management Review, vol. 1, Nov. 2008, doi: 10.2991/itmr.2008.1.2.4.
[31] E. Carmel, Y. Dubinsky, and A. Espinosa, "Follow The Sun Software Development: New Perspectives, Conceptual Foundation, and Exploratory Field Study," in 2009 42nd Hawaii International Conference on System Sciences, 2009, pp. 1–9. doi: 10.1109/HICSS.2009.218.
[32] E. Carmel, J. Espinosa, and Y. Dubinsky, "'Follow the Sun' Workflow in Global Software Development," J. of Management Information Systems, vol. 27, pp. 17–38, Jul. 2010, doi: 10.2753/MIS0742-1222270102.
[33] A. Ziegler et al., Productivity Assessment of Neural Code Completion. 2022. [Online]. Available: https://arxiv.org/abs/2205.06537.
[34] M. Chen et al., Evaluating Large Language Models Trained on Code. 2021. [Online]. Available: https://arxiv.org/abs/2107.03374
[35] F. F. Xu, U. Alon, G. Neubig, and V. J. Hellendoorn, "A systematic evaluation of large language models of code," in Proceedings of the 6th ACM SIGPLAN International Symposium on Machine Programming, 2022, pp. 1–10. doi: 10.1145/3520312.3534862.
[36] M. Wessel, I. Steinmacher, I. Wiese, and M. A. Gerosa, "Should I Stale or Should I Close? An Analysis of a Bot That Closes Abandoned Issues and Pull Requests," in 2019 IEEE/ACM 1st International Workshop on Bots in Software Engineering (BotSE), 2019, pp. 38–42. doi: 10.1109/BotSE.2019.00018.
[37] International Association of Project Managers, "The V-model Explained," iapm.net. https://www.iapm.net/en/blog/v-model/ (accessed June 28, 2024).
[38] A. Nüßgen et al., "Leveraging Robust Artificial Intelligence for Mechatronic Product Development—A Literature Review." International Journal of Intelligence Science, 14, 1-21. 2024, https://doi.org/10.4236/ijis.2024.141001
[39] "Recap: The V-Model," aiotplaybook.org. https://aiotplaybook.org/index.php?title=Agile_V-Model (accessed June 28, 2024).
[40] R. York and J. Mcgee, "Understanding the Jevons paradox," Environmental Sociology, vol. 2, pp. 1–11, Dec. 2015, doi: 10.1080/23251042.2015.1106060.

APPENDIX

TABLE I
AI INTEGRATION POINTS THROUGHOUT THE V-MODEL

| Phase | Activity | Example areas of AI use |
|---|---|---|
| Verification | Requirements Gathering | Requirements Capturing (NLP, Cv, Etc.), Automated Requirements Generation, Requirements Clustering, Requirements Indexing and Matching, Stakeholder Sentiment Analysis, Requirement Conflict Detection, Automated Traceability Matrix Generation |
| | System Analysis | System Modeling, Risk Analysis And Mitigation Suggestions, Feasibility Assessment, Resource Estimation And Allocation, Dependency Mapping And Analysis, Compliance And Regulatory Requirement Checking |
| | Software Design | Generative Ui/Ux Design, Generative UML + Data Modeling, Architecture Pattern Recommendation, Microservices Design Optimization, Api Design Generation |
| | Module Design | Module Decomposition, Interface Design Optimization, Code Reusability Analysis, Performance Optimization Suggestions, Security Recommendations |
| Coding | Code Generation | Code generation, code optimization, refactoring suggestions, documentation generation, legacy code analysis |
| Validation | Unit Testing | Test case generation, test data creation, mutation testing, code coverage analysis, performance benchmarking, static code analysis |
| | Integration Testing | Integration test scenario generation, API testing automation, continuous integration optimization, regression test selection |
| | System Testing | End-to-end test scenario generation, load and stress test design, security vulnerability scanning, user scenario simulation, system performance optimization |
| | Acceptance Testing | Test generation, adaptive test management, predictive system error forecasting, test data creation, user acceptance criteria verification, automated user experience testing |